\documentclass{article}

\usepackage{times}
\usepackage{graphicx} 

\usepackage{subfigure} 

\usepackage{natbib}

\usepackage{algorithm}
\usepackage{algorithmic}

\usepackage{hyperref}

\usepackage[accepted]{icml2015} 

\usepackage{array}  
\newcolumntype{C}[1]{>{\centering\let\newline\\\arraybackslash\hspace{0pt}}m{#1}}
\usepackage{comment}

\icmltitlerunning{Understanding Music Playlists}

\begin{document} 

\twocolumn[
\icmltitle{Understanding Music Playlists}
\icmlauthor{Keunwoo Choi}{keunwoo.choi@qmul.ac.uk}
\icmladdress{Queen Mary University of London,
            E1 4NS, London, The United Kingdom}
\icmlauthor{Gy\"{o}rgy Fazekas, Mark Sandler}{g.fazekas@qmul.ac.uk, m.sandler@qmul.ac.uk}
\icmladdress{Queen Mary University of London,
            E1 4NS, London, The United Kingdom}

\icmlkeywords{music recommendation, playlist}

\vskip 0.3in
]


\begin{abstract} 
As music streaming services dominate the music industry, the playlist is becoming an increasingly crucial element of  music consumption. Consequently, the music recommendation problem is often casted as a playlist generation problem. Better understanding of the playlists is therefore necessary for developing better playlist generation algorithms. In this work, we analyse two playlist datasets to investigate some commonly assumed hypotheses about playlists. Our findings indicate that deeper understanding of playlists is needed to provide better prior information and improve machine learning algorithms in the design of recommendation systems.

\end{abstract}

\section{Introduction and Backgrounds}
\label{intro}
As streaming services become more popular, more people consume music by listening to playlists. People are now looking for a \textit{good} playlist filled with \textit{right songs} at \textit{right positions}. Therefore, it is natural that the focus of music recommendation problem is shifting towards the playlist generation problem.

However, the understanding of playlists is still shallow. Deeper understanding of playlists is necessary to improve playlist generation algorithms. In this study, we present an investigation of playlist features derived from audio features and metadata. Our findings provide insight which may be useful in the design of music recommendation system.





Firstly, the definition of a playlist becomes clear by comparing it to the \textit{mix}. Although both are sequences of songs, a playlist can be described as an artefact for \textit{personal use with less strictly defined theme}, while a mix as a \textit{set of music to compile a custom CD with a strongly defined theme} for DJing purpose \cite{cunningham2006more}. The personal-use aspect of playlists suit well to the more recent concept of playlists. 

Additional constraints have been proposed to model a playlist. One of the most common one is \textit{global similarity} e.g. \cite{platt2001learning}, \cite{ragno2005inferring} defining a playlist as \textit{a sequence of songs that are similar to each other}.
Another common constraint is on the relationship of two consecutive songs.
In ~\cite{mcfee2012hypergraph}, a bi-gram language model that treat songs as words was used with first-order Markov process. More generalised constraints are introduced in ~\cite{pauws2006fast}, where a playlist is loosely defined as \textit{a sequence of the right song at the right positions.} This may be formalised by categorised constraints into three types; \textit{Unary}, \textit{binary}, and \textit{global}. 

However, constraints are not generally agreed upon yet. For example, how an algorithm compromises between similarity and diversity? Does each constrain depend on the user, the audio features, and/or label, if so, how?
There is no doubt that answering these questions will lead to improvement in the  design of machine learning algorithms in recommendation systems; for example, by informing a correct segmentation of data and corresponding priors, or providing the design of complex multimodal algorithms.


\section{Playlist Data Analysis}
\label{sec:analysis}

In this work, we use two datasets; a playlist dataset that we crawled directly from \textit{Deezer} (named as \textit{Deezer-2015}) and \textit{AoTM-2011} dataset, which is introduced in ~\cite{mcfee2012hypergraph}. Both of them have more than 50K playlists with roughly 100K songs. 

There are two main differences between the two datasets; First, only AoTM-2011 is labeled with category names including genre (e.g. \textit{Rock}, \textit{Jazz}), mood (e.g. \textit{Depression}, \textit{Romantic}), or activity (e.g. \textit{Road Trip}). Second, the song features of Deezer-2015 are high-level audio features and metadata from EchoNest, while that of AoTM-2011 includes low-level audio timbre features.

\par
\subsection{Does clusters exist in audio-based feature domain?}
According to ~\cite{jennings2007net}, music listeners are categorised in four types; \textit{Savant}, \textit{enthusiast}, \textit{causal}, and \textit{indifferent}. The hypothesis here is whether we can observe multiple distinct clusters in the feature domain, possibly due to different playlist structures by each user groups. We used both datasets to verify this hypothesis.
We found that the distributions of all the single features showed uni-modality. We also investigated its distribution after reducing the dimensions of the feature space into three using Principal Component Analysis. However, the observed distribution was still uni-modal. 

As a result, the content-based song features was ineffective to discriminate different user groups. We suggest that user analysis should be done with relevant user data - demographics, psychographic, and user behaviours. 

\let\thefootnote\relax\footnotetext{This paper has been supported by EPSRC Grant EP/L019981/1, Fusing Audio and Semantic Technologies for Intelligent Music Production and Consumption
}%

\subsection{Similarity vs. Diversity}
A good playlist is supposed to be filled with similar songs, but not \textit{too} similar, as in \cite{cunningham2006more}. In other words, both \textit{similarity} and \textit{diversity} are expected to make a good playlist. To verify this, we calculated the average cosine distances between feature vectors extracted from songs in all playlists and compare its distribution with the distribution of randomly picked 5,000 pairs of the songs. AoTM-2011 is used in this experiment.

The result verifies the hypothesis well; though their averages were similar (0.080 vs. 0.095), the 25-75 percentile are [0.061, 0.094] vs. [0.041, 0.126], in within-playilist and arbitrary pairs, respectively. This shows there may exist \textit{proper} range of similarity that constitutes a good playlist.

\subsection{Does playlist similarity differs by category?}

\begin{figure}[ht]
\begin{center}
\centerline{\includegraphics[width=\columnwidth]{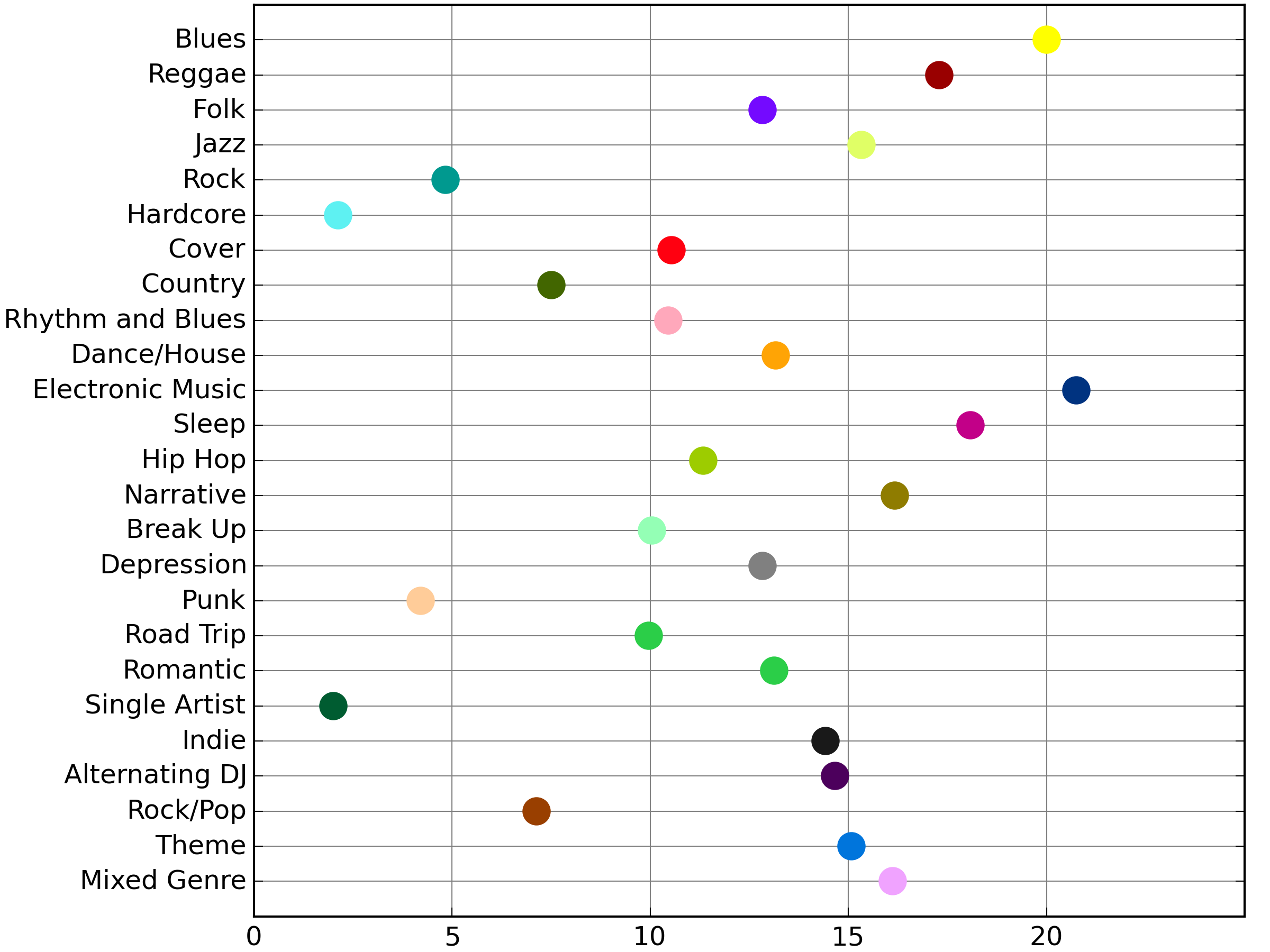}}
\vskip -0.1in
\caption{Average rankings of the variances of playlist features. Higher raking indicates smaller variance.}
\label{figure:rankings}
\end{center}
\vskip -0.48in
\end{figure}

In ~\cite{mcfee2012hypergraph}, category-specific models are compared to a global model, where category models show better result on modelling the playlists except \textit{Narrative} and \textit{Rock} (tested against AoTM-2011). Not surprisingly, \textit{Mixed} category shows the lowest similarity between songs in the playlist.

We calculated the variances of playlist features in each categories to investigate the similarity of audio features using AoTM-2011. The variance should be smaller if the playlists in the category are similar than those in other categories, therefore the \textit{Mixed} category can be used as an \textit{anchor}. Note that the consistency is measured at a \textit{playlist} level, not a song level.

As shown in the Figure \ref{figure:rankings}, \textit{Single Artist}, \textit{Hardcore}, and \textit{Punk} show the smallest variances, followed by \textit{Rock}, \textit{Rock/Pop}, and \textit{Country}. It is generally agreeable since labels such as \textit{punk}, \textit{hardcore}, \textit{rock}, and \textit{single artist} are well characterised by timbre. On the other hand, 5 categories out of 25 show larger variances than the variance of \textit{Mixed}; they are \textit{Sleep}, \textit{Electronic Music}, \textit{Blues}, \textit{Raggae}, and \textit{Blues}. This may be interpreted, for example, that the audio features of \textit{Jazz} varies due to the variety of instrumentations. More importantly, however, it shows that when generating playlists, different levels of similarity is expected for different genre or context labels. 



\section{Conclusion}
This work details the investigation of three hypotheses about music playlists. First, clusters that may exist due to different user groups are not observed in audio and metadata feature domain, which suggests that a user-model (for user-segmentation) is recommended to be built based on user information and behaviours. Second, the observed variance in song features differs for different playlist categories suggesting there is a difference in the optimal similarity between songs in different situations, rather than using the most similar songs. Lastly, audio feature consistency is shown to be different by the categories. It indicates that the parameter for within-playlist similarity should be learned separately when optimising playlist generation algorithms. Our findings may lead to better priors in the design of machine learning algorithms for music recommendation systems.



\bibliography{example_paper}

\begin{thebibliography}{6}
\providecommand{\natexlab}[1]{#1}
\providecommand{\url}[1]{\texttt{#1}}
\expandafter\ifx\csname urlstyle\endcsname\relax
  \providecommand{\doi}[1]{doi: #1}\else
  \providecommand{\doi}{doi: \begingroup \urlstyle{rm}\Url}\fi

\bibitem[Cunningham et~al.(2006)Cunningham, Bainbridge, and
  Falconer]{cunningham2006more}
Cunningham, Sally~Jo, Bainbridge, David, and Falconer, Annette.
\newblock "more of an art than a science": Supporting the creation of playlists
  and mixes.
\newblock \emph{ISMIR}, 2006.

\bibitem[Jennings(2007)]{jennings2007net}
Jennings, David.
\newblock \emph{Net, blogs and rock'n'roll: how digital discovery works and
  what it means for consumers, creators and culture}.
\newblock Nicholas Brealey Publishing, 2007.

\bibitem[McFee \& Lanckriet(2012)McFee and Lanckriet]{mcfee2012hypergraph}
McFee, Brian and Lanckriet, Gert~RG.
\newblock Hypergraph models of playlist dialects.
\newblock In \emph{ISMIR}, pp.\  343--348, 2012.

\bibitem[Pauws et~al.(2006)Pauws, Verhaegh, and Vossen]{pauws2006fast}
Pauws, Steffen, Verhaegh, Wim, and Vossen, Mark.
\newblock Fast generation of optimal music playlists using local search.
\newblock In \emph{ISMIR}, 2006.

\bibitem[Platt et~al.(2001)Platt, Burges, Swenson, Weare, and
  Zheng]{platt2001learning}
Platt, John~C, Burges, Christopher~JC, Swenson, Steven, Weare, Christopher, and
  Zheng, Alice.
\newblock Learning a gaussian process prior for automatically generating music
  playlists.
\newblock In \emph{NIPS}, pp.\  1425--1432, 2001.

\bibitem[Ragno et~al.(2005)Ragno, Burges, and Herley]{ragno2005inferring}
Ragno, Robert, Burges, Christopher~JC, and Herley, Cormac.
\newblock Inferring similarity between music objects with application to
  playlist generation.
\newblock In \emph{Proceedings of the 7th ACM SIGMM international workshop on
  Multimedia information retrieval}, pp.\  73--80. ACM, 2005.

\end{thebibliography}
\bibliographystyle{icml2015}

\end{document}